\documentclass[aps,prl,twocolumn,superscriptaddress,floatfix]{revtex4-1}
\pdfoutput=1

\usepackage{pdfpages}
\usepackage{graphicx,graphics}
\usepackage{dcolumn}
\usepackage{amsmath,amssymb,amsfonts}
\usepackage{latexsym,verbatim}
\usepackage{bm}
\usepackage{color}
\usepackage{ulem}
\usepackage[percent]{overpic}
\usepackage[breaklinks=true,colorlinks,citecolor=blue,linkcolor=blue,urlcolor=blue]{hyperref}

\begin{document}

\title{Accessing phonon polaritons in hyperbolic crystals by ARPES}

\author{Andrea Tomadin}
\affiliation{NEST, Istituto Nanoscienze-CNR and Scuola Normale Superiore, I-56126 Pisa,~Italy}

\author{Alessandro Principi}
\affiliation{Department of Physics and Astronomy, University of Missouri, Columbia, Missouri 65211,~USA}

\author{Justin C.W. Song}
\affiliation{Walter Burke Institute for Theoretical Physics and Institute for Quantum Information and Matter, California Institute of Technology, Pasadena, CA 91125, USA} 

\author{Leonid S. Levitov}
\email{levitov@mit.edu}
\affiliation{Department of Physics, Massachusetts Institute of Technology, Cambridge, Massachusetts 02139,~USA} 

\author{Marco Polini}
\email{marco.polini@icloud.com}
\affiliation{NEST, Istituto Nanoscienze-CNR and Scuola Normale Superiore, I-56126 Pisa,~Italy}
\affiliation{Istituto Italiano di Tecnologia, Graphene Labs, Via Morego 30, I-16163 Genova,~Italy}

\begin{abstract}
Recently studied hyperbolic materials host unique phonon-polariton (PP) modes. 
The ultra-short wavelengths of these modes, which can be much smaller than those of conventional exciton-polaritons, are of high interest for extreme sub-diffraction nanophotonics schemes. 
Polar hyperbolic materials such as hexagonal boron nitride can be used to realize strong long-range coupling between PP modes and extraneous charge degrees of freedom. 
The latter, in turn, can be used to control and probe PP modes. 
Of special interest is coupling between PP modes and plasmons in an adjacent graphene sheet, which opens the door to accessing PP modes by angle-resolved photoemission spectroscopy (ARPES). 
A rich structure in the graphene ARPES spectrum due to PP modes is predicted, providing a new probe of PP modes and their coupling to graphene plasmons.
\end{abstract}

\maketitle

\noindent {\it Introduction.---}The intrinsic hyperbolic character~\cite{hyperbolicmaterials} of hexagonal boron nitride (hBN) grants a unique platform for realizing deep-subwavelength nanophotonic schemes. 
Key to these developments are phonon-polariton (PP) modes that exist within {\it reststrahlen} frequency bands~\cite{dai_science_2014,caldwell_naturecommun_2014}, characterized by wavelengths that can be as small as $1$-$100~{\rm nm}$.
Highly directional, these modes exhibit deep sub-diffraction confinement of light with wavelengths far shorter than those of exciton-polaritons in semiconductor microcavities~\cite{byrnes_naturephys_2014}.
PPs have been shown to propagate with low losses~\cite{dai_science_2014,caldwell_naturecommun_2014} besting artificial metallic-resonator metamaterial schemes, and opening the door to hyperlensing~\cite{li_arxiv_2015,dai_arxiv_2015_II}.

\begin{figure}[h!]
\includegraphics[width=0.95\linewidth]{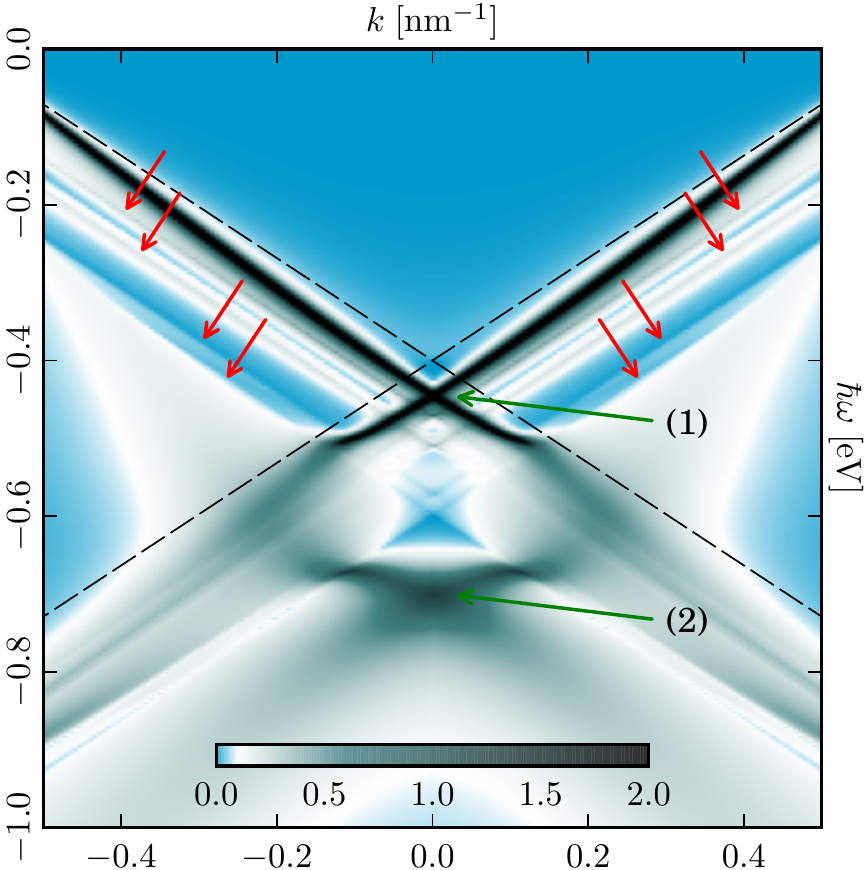}
\caption{\label{fig:one}
(Color online) Signatures of PP modes in the quasiparticle spectral function ${\cal A}({\bm k},\omega)$ of a doped graphene sheet placed over a hBN slab, obtained from Eqs.~(\ref{eq:imsigma}) and~(\ref{eq:spectralfunction}).
Note the black linearly-dispersing quasiparticle bands, which display a clear Dirac crossing labeled by (1), and the  broad spectral feature labeled by (2) due to the emission of the plasmon-phonon polariton mode with highest energy in Fig.~\ref{fig:two}(a).
The Fermi energy is positioned at $\omega = 0$.
Emission of polariton modes [see Fig.~\ref{fig:two}(a)] by the holes created by photo-excited electrons gives rise to four dispersive satellite bands running parallel to the main quasiparticle bands (marked by red arrows). 
The feature (2) is mainly plasmonic,whereas the satellite bands, crossing at $k = 0$ between features (1) and (2), are entirely due to Fabry-P\'erot hBN phonon-polariton modes.
Parameters used: Fermi energy $\varepsilon_{\rm F} = 400~{\rm meV}$,  hBN slab thickness $d = 60~{\rm nm}$,  $\epsilon_{\rm a} = 1$ (vacuum), $\epsilon_{\rm b} = 3.9$ (${\rm SiO}_2$).
The colorbar refers to the values of $\hbar{\cal A}({\bm k},\omega)$ in ${\rm eV}$.}
\end{figure}

Key to harnessing PP modes is gaining access to their response over a wide wavenumber and energy bandwidths.
However, to date PPs have only been studied within a small frequency range limited by laser choice (e.g.~ $167~{\rm meV} \lesssim \hbar \omega \lesssim 198~{\rm meV}$ via scattering-type near-field optical spectroscopy technique~\cite{dai_science_2014}), or at specific wavelengths fixed by the sample geometry via Fourier transform infrared spectroscopy of nanofabricated nanopillars~\cite{caldwell_naturecommun_2014}.
New approaches allowing to resolve the PP modes at shorter wavelengths and over a broad range of energies are therefore highly desirable.

Here we describe an angle-resolved photoemission spectroscopy (ARPES)~\cite{damascelli_rmp_2003} scheme to achieve broadband energy-resolved access to ultra-short wavelength PPs in hBN.
At first glance, ARPES access to PPs in a wide-bandgap insulator (hBN) where no free carriers are available may seem counterintuitive.
However, the key to our protocol lies in coupling PPs to charge degrees of freedom in a conductor (e.g.~graphene) placed nearby the hyperbolic crystal of interest (hBN), prepared in a slab geometry.
Strong coupling~\cite{amorim_prb_2012,brar_nanolett_2014,principi_prb_2014,woessner_naturemater_2015,dai_arxiv_2015_I,kumar_arxiv_2015} between hBN Fabry-P\'erot PP modes and the collective charge oscillations (i.e.~Dirac plasmons~\cite{grapheneplasmons}) in a doped graphene sheet placed over a hBN slab gives rise to new channels for quasiparticle decay yielding a rich structure of dispersive satellite features---marked by red arrows in Fig.~\ref{fig:one}---in the graphene ARPES spectrum ${\cal A}({\bm k}, \omega)$.
Since hBN Fabry-P\'erot PPs are controlled by slab thickness, the composite G/hBN structure features a novel ARPES spectrum with features that are highly tunable by thickness.

The greatest practical advantage of this approach is that ARPES achieves extreme resolution over a wide range of wave vectors ${\bm k}$ (from the corners $K, K^\prime$ of the graphene Brillouin zone to the Fermi wave number $k_{\rm F}$ in graphene) and energies $\hbar\omega$, with all energies below the Fermi energy being probed simultaneously.
This gives an additonal benefit, besides tunability, in that the entire range of frequencies and wavenumbers can be covered within a single experiment.
It is remarkable that a one-atom-thick conducting material like graphene, once placed over an insulating hyperbolic crystal, enables ARPES studies of PP modes over the full range of wave vectors and energies of interest.

From a more fundamental perspective, ARPES will also be an ideal tool to investigate whether effective electron-electron interactions mediated by the exchange of PPs are capable of driving electronic systems towards correlated states. 
Finally, looking at our results from the point of view of graphene optoelectronics, one can envision situations in which the tunable coupling between graphene quasiparticles and the complex excitations of its supporting substrate can be used to achieve control over the {\it spectral} properties of graphene carriers, including their decay rates, renormalized velocities, etc.
This degree of tunability may have important implications on the performance of graphene-based photodetectors~\cite{koppens_naturenano_2014}. 

\begin{figure}
\begin{overpic}[width=\linewidth]{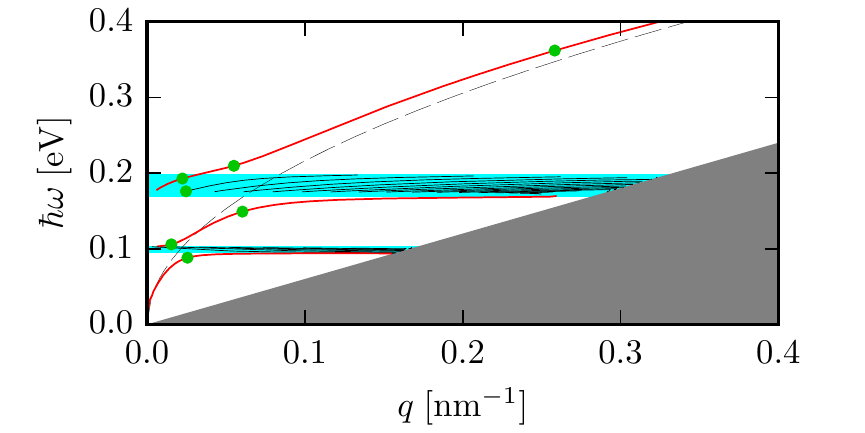}\put(2,50){(a)}\end{overpic}
\begin{overpic}[width=\linewidth]{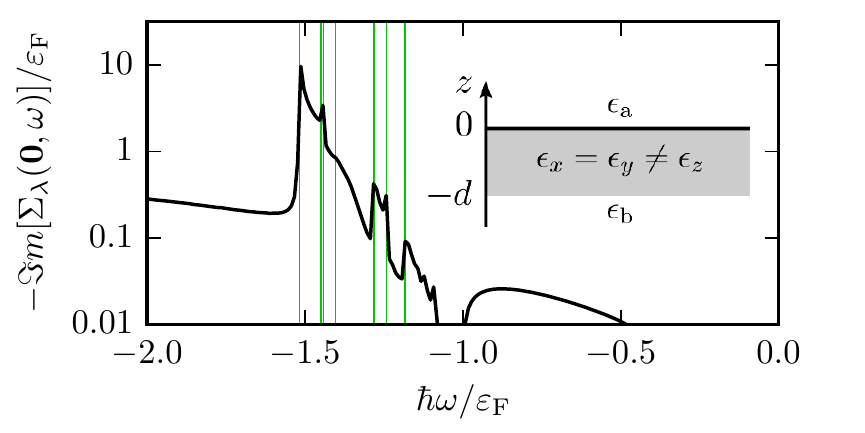}\put(2,50){(b)}\end{overpic}
\begin{overpic}[width=\linewidth]{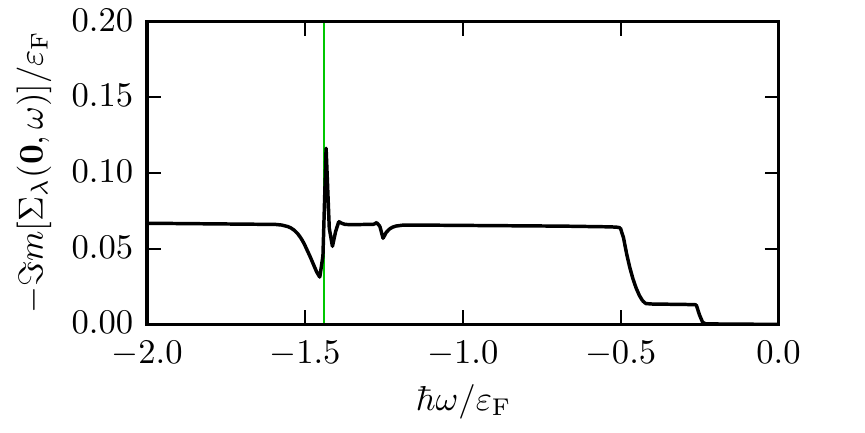}\put(2,50){(c)}\end{overpic}
\caption{\label{fig:two}
(Color online) Panel (a) Dispersion relation $\Omega_{\bm q}$ of hybrid plasmon-phonon polaritons (solid lines) with parameters as in Fig.~\ref{fig:one}.
The dashed line represents the dispersion relation of a Dirac plasmon~\cite{grapheneplasmons} in graphene, in the absence of hBN  phonons.
Horizontal cyan areas denote the hBN reststrahlen bands.
The grey-shaded area represents the intra-band particle-hole continuum in graphene.
Green filled circles represent the points where the plasmon-phonon polariton group velocity equals the graphene Fermi velocity $v_{\rm F}$.
Panel (b) The quantity $-{\rm Im}[\Sigma_{\lambda}({\bm k}, \omega)]$ (in units of $\varepsilon_{\rm F}$ and evaluated at ${\bm k} = {\bm 0}$) is shown as a function of the rescaled frequency $\hbar\omega/\varepsilon_{\rm F}$.
Green vertical lines denote the values of $\hbar \omega/\varepsilon_{\rm F}$ at which a plasmon-phonon polariton peak is expected.
The vertical axis is in logarithmic scale.
The inset shows a side view of the vertical heterostructure analyzed in this work.
Panel (c) Same as in panel (b) but in the absence of dynamical screening due to electron-electron interactions in graphene: these numerical results have been obtained by replacing $W_{{\bm q}, \omega} \to V_{{\bm q}, \omega}$ in Eq.~(\ref{eq:imsigma}).
A polaron peak is clearly visible.}
\end{figure}

\noindent {\it Phonon and plasmon-phonon polaritons.---}We consider a vertical heterostructure---see inset in Fig.~\ref{fig:two}(b)---composed of a graphene sheet located at $z = 0$ and placed over a homogeneous anisotropic insulator of thickness $d$ with dielectric tensor $\hat{\bm \epsilon} = {\rm diag}(\epsilon_{x},\epsilon_{y},\epsilon_{z})$.
Homogeneous and isotropic insulators with dielectric constants $\epsilon_{\rm a}$ and $\epsilon_{\rm b}$ fill the two half-spaces $z>0$ and $z < -d$, respectively. 
The Fourier transform $V_{{\bm q}, \omega}$ of the Coulomb interaction potential, as dressed by the presence of a {\it uniaxial} ($\epsilon_{y} = \epsilon_{x}$) dielectric, is given by 
\begin{equation}\label{eq:effective_interaction_uniaxial}
V_{{\bm q}, \omega} = \varphi_{\bm q}\frac{\sqrt{\epsilon_{x} \epsilon_{z}} +\epsilon_{\rm b}
\tanh(qd\sqrt{\epsilon_{x}/\epsilon_{z}})}{\sqrt{\epsilon_{x} \epsilon_{z}} + (\epsilon_{x} \epsilon_{z} + \epsilon_{\rm b} \epsilon_{\rm a})\tanh(qd\sqrt{\epsilon_{x}/\epsilon_{z}})/(2{\bar \epsilon})}, 
\end{equation}
where $v_{\bm q} = 2 \pi e^2 /(q \bar{\epsilon})$ with $\bar{\epsilon} = (\epsilon_{\rm a} + \epsilon_{\rm b})/2$ is the ordinary 2D Coulomb interaction potential.
A more general equation, which is also valid in the case $\epsilon_{y} \neq \epsilon_{x}$, can be found in Sect.~I of Ref.~\onlinecite{SMF}.

In the case of hBN, the components $\epsilon_{x}$ and $\epsilon_{z}$ of the dielectric tensor have an important dependence on frequency $\omega$ in the mid infrared~\cite{geick_pr_1966}.
The simplest parametrization formulas for $\epsilon_{x, z} = \epsilon_{x, z}(\omega)$ are reported in Sect.~I of Ref.~\onlinecite{SMF} and have been used for the numerical calculations.
More realistic parametrizations can be found in the Supplementary Information of Ref.~\onlinecite{woessner_naturemater_2015}.

Standing PP modes~\cite{dai_science_2014} correspond to poles of the dressed interaction $V_{{\bm q}, \omega}$ {\it inside} the reststrahlen bands.
These can be found by looking at the zeroes of the denominator in Eq.~(\ref{eq:effective_interaction_uniaxial}), $\sqrt{|\epsilon_{x}(\omega) \epsilon_{z}(\omega)|} + (2{\bar \epsilon})^{-1}[\epsilon_{x}(\omega) \epsilon_{z}(\omega) + \epsilon_{\rm b} \epsilon_{\rm a}]\tan[qd\sqrt{|\epsilon_{x}(\omega)/\epsilon_{z}(\omega)|}]=0$.
Illustrative numerical results for $d =10~{\rm nm}$ and $d= 60~{\rm nm}$ are reported in Fig.~1 of Ref.~\onlinecite{SMF}.
Analytical expressions, which are valid for $qd \ll 1$ and $qd\gg1$, are available~\cite{SMF} in the case in which phonon losses in hBN are neglected.
For sufficiently thick hBN slabs, there can be modes with group velocity equal to the graphene Fermi velocity $v_{\rm F}$.

Standing PP modes in a hBN slab couple to Dirac plasmons in a nearby graphene sheet.
Such coupling is captured by the random phase approximation (RPA)~\cite{Giuliani_and_Vignale}.
In the RPA, one introduces the dynamically screened interaction
\begin{equation}\label{eq:screenedinteraction}
W_{{\bm q}, \omega} = \frac{V_{{\bm q}, \omega}}{\varepsilon({\bm q},\omega)} \equiv \frac{V_{{\bm q}, \omega}}{1 - V_{{\bm q},\omega}\chi_{0}(q,\omega)}.
\end{equation}
Here $\varepsilon({\bm q}, \omega)$ is the RPA dielectric function and $\chi_0(q,\omega)$ is the 
density-density response function of a 2D massless Dirac fermion fluid~\cite{polarizationfunction}.
While the poles of $V_{{\bm q}, \omega}$ physically yield slab PP modes, new poles of $W_{{\bm q}, \omega}$ emerge from electron-phonon interactions.
These are weakly-damped solutions $\omega = \Omega_{\bm q} - i 0^+$ of the equation $\varepsilon({\bm q}, \omega)=0$.
We have solved this equation numerically and illustrative results for $\varepsilon_{\rm F} = 400~{\rm meV}$ 
and $d =60~{\rm nm}$ are shown in Fig.~\ref{fig:two}(a).
(Results for different values of $\varepsilon_{\rm F}$ and $d$ can be found in Sect.~III of Ref.~\onlinecite{SMF}.)
Solid lines represent plasmon-phonon polaritons that emerge from the hybridization between the Dirac plasmon~\cite{grapheneplasmons} in graphene (dashed line) and standing PP waves in the hBN slab.
The solid red lines denote three polariton branches with a strong degree of plasmon-phonon hybridization.
On the contrary, black solid lines denote practically unhybridized slab PP modes. 
We clearly see that there are several plasmon-phonon polariton modes (green circles) with group velocity equal to $v_{\rm F}$.
These modes couple strongly to quasiparticles in graphene, as we now proceed to demonstrate.

\noindent{\it Quasiparticle decay rates.---}An excited quasiparticle with momentum ${\bm k}$ and energy $\hbar\omega$, created in graphene in an ARPES experiment~\cite{bostwick_naturephys_2007,zhou_naturemater_2007,bostwick_science_2010,walter_prb_2011,siegel_pnas_2011}, can decay by scattering against the excitations of the Fermi sea, i.e.~electron-hole pairs and collective modes.
The decay rate $\hbar/\tau_\lambda({\bm k},\omega)$ for these processes can be calculated~\cite{Giuliani_and_Vignale} from the imaginary part of the retarded quasiparticle self-energy $\Sigma_\lambda({\bm k},\omega)$, i.e.~$\hbar/\tau_\lambda({\bm k}, \omega) =  - 2{\rm Im}\left[ \Sigma_\lambda({\bm k},\omega)\right]$.
In the RPA and at zero temperature we have~\cite{polini_prb_2008,hwang_prb_2008}
\begin{eqnarray}\label{eq:imsigma}
{\rm Im}\left[ \Sigma_\lambda({\bm k},\omega)\right]&=&\sum_{\lambda'}
\int \frac{d^2{\bm q}}{(2\pi)^2} {\rm Im} \left[W_{{\bm q}, \omega -\xi_{\lambda^\prime, {\bm k} + {\bm q}}}\right]{\cal F}_{\lambda\lambda'}\nonumber\\
&\times&\left[\Theta(\hbar\omega-\xi_{\lambda', {\bm k} + {\bm q}}) - \Theta(-\xi_{\lambda^\prime, {\bm k} + {\bm q}})\right].
\end{eqnarray}
Here ${\cal F}_{\lambda\lambda'} \equiv [1+\lambda \lambda^\prime\cos{(\theta_{{\bm k}, {\bm k} + {\bm q}})}]/2$ is the chirality factor~\cite{polini_prb_2008,hwang_prb_2008}, $\xi_{\lambda, {\bm k}} = \lambda \hbar v_{\rm F} k - \varepsilon_{\rm F}$ is the Dirac band energy measured from the Fermi energy $\varepsilon_{\rm F}$ ($\lambda, \lambda^\prime = \pm 1$), and $\Theta(x)$ is the usual Heaviside step function.
The quantity $\hbar\omega$ is also measured from the Fermi energy and, finally, $\theta_{{\bm k}, {\bm k} + {\bm q}}$ is the angle between ${\bm k}$ and ${\bm k} + {\bm q}$. 
Eq.~(\ref{eq:imsigma}) reduces to the standard Fermi golden rule when only terms of ${\cal O}(V^2_{{\bm q}, \omega})$ are retained.
Physically, it describes the decay rate of a process in which an initial state with momentum ${\bm k}$ and energy $\hbar\omega$ (measured from $\varepsilon_{\rm F}$) decays into a final state with momentum ${\bm k} +{\bm q}$ and energy $\xi_{\lambda', {\bm k} + {\bm q}}$ (measured from $\varepsilon_{\rm F}$).
For $\omega < 0$, the self-energy expresses the decay of {\it holes} created inside the Fermi sea, which scatter to a final state, by exciting the Fermi sea.
Fermi statistics requires the final state to be {\it occupied} so both band indices $\lambda' = \pm 1$ are allowed in the case $\varepsilon_{\rm F} > 0$ that we consider here.
Since ARPES measures the properties of holes produced in the Fermi sea by photo-ejection, only $\omega < 0$ is relevant for this experimental probe in an $n$-doped graphene sheet. 

It is convenient to discuss the main physical features of ${\rm Im}\left[ \Sigma_\lambda({\bm k},\omega)\right]$ for an initial hole state with momentum ${\bm k} = {\bm 0}$.
In this case, the 2D integral in Eq.~(\ref{eq:imsigma}) reduces to a simple 1D quadrature.
The initial hole energy is $E_{\rm i} = \hbar\omega + \varepsilon_{\rm F}$.
The final hole energy is $E_{\rm f} = \xi_{\lambda', {\bm q}} +\varepsilon_{\rm F} = \lambda' \hbar v_{\rm F} q$.
When the difference $\Delta_{\lambda',q} \equiv E_{\rm f} - E_{\rm i}$ is equal to the real part of the mode energy $\hbar\Omega_{\bm q}$, the initial hole, which has been left behind after the photo-ejection of an electron, can decay by emitting a plasmon-phonon polariton.
Since $\hbar\Omega_{\bm q} > \hbar v_{\rm F} q$, but $\Delta_{\lambda',q} \leq \hbar v_{\rm F} q$ for intraband transitions,
an initial hole state with $E_{\rm i}<0$ (i.e.~initial hole state in valence band) can decay only into a final hole state with $E_{\rm f} >0$ (i.e.~final hole state in conduction band).
In particular, when $d \Omega_{\bm q}/dq = \hbar^{-1} d \Delta_{\lambda',q}/dq = \lambda' v_{\rm F}$, such decay process is {\it resonant}.
When these conditions are met, the inter-band contribution to ${\rm Im} \left[ \Sigma_\lambda({\bm 0},\omega)\right]$ peaks at a characteristic value of $\omega$ and the Kramers-Kronig transform ${\rm Re}[\Sigma_\lambda({\bm 0},\omega)]$ changes sign rapidly around that frequency.
Within RPA, a satellite quasiparticle emerges~\cite{classics}, which is composed by a hole that moves with the same speed of a plasmon-phonon polariton.
This is a solution of the Dyson equation, distinct from the ordinary quasiparticle solution that becomes the Landau pole of the one-body Green's function as $k \to k_{\rm F}$ and $\omega \to 0$.
\begin{figure}[t]
\begin{overpic}[width=\linewidth]{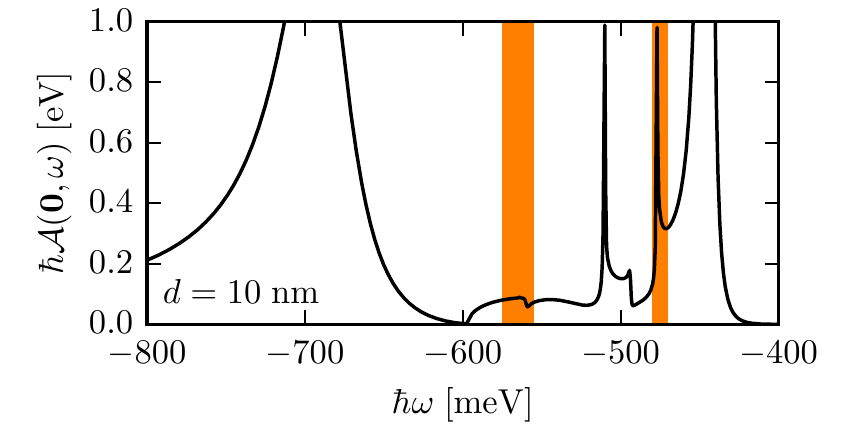}\put(2,50){(a)}\end{overpic}
\begin{overpic}[width=\linewidth]{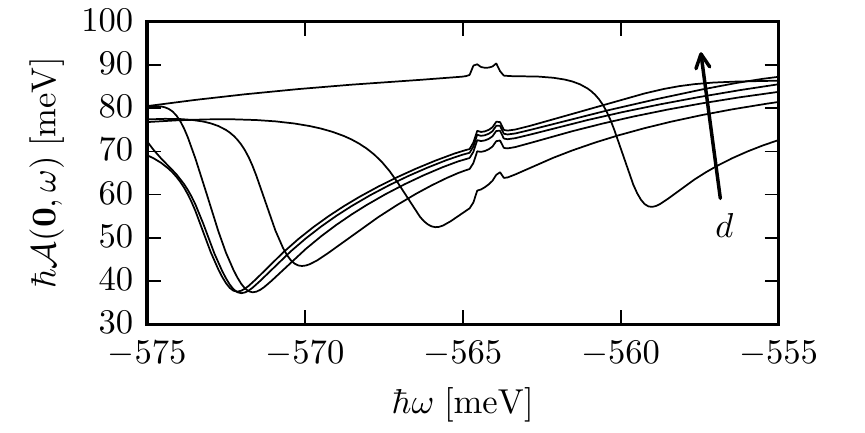}\put(2,50){(b)}\end{overpic}
\begin{overpic}[width=\linewidth]{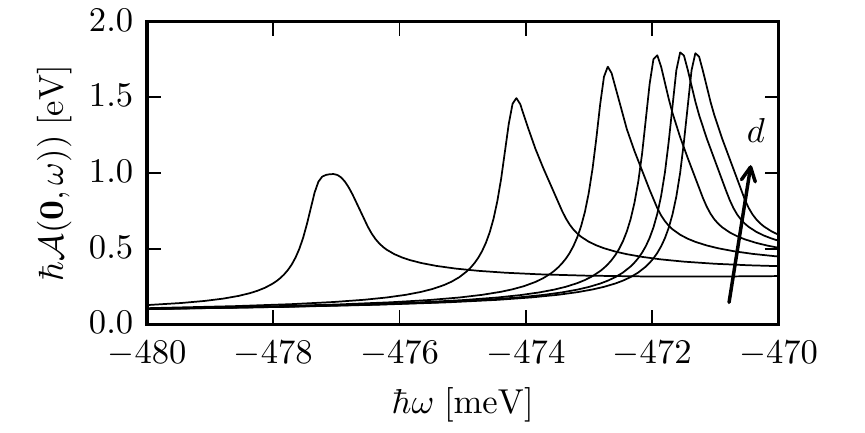}\put(2,50){(c)}\end{overpic}
\caption{\label{fig:three}
(Color online) Panel (a) The quasiparticle spectral function ${\cal A}({\bm k},\omega)$ evaluated at ${\bm k} = {\bm 0}$ ($|{\bm k}| = 10^{-3}~k_{\rm F}$ has been used in the numerical calculations) is plotted as a function of $\hbar \omega$.
This plot refers to $d = 10~{\rm nm}$.
The other parameters are as in Fig.~\ref{fig:one}.
Panels (b) and (c) Dependence on the hBN slab thickness $d$ of the spectral function features highlighted by vertical orange-shaded regions in panel (a). 
Different curves correspond to values of $d$ on a uniform mesh from $d = 10~{\rm nm}$ to $d = 60~{\rm nm}$.
Arrows indicate how spectral features evolve by increasing $d$.}
\end{figure}

The quantity ${\rm Im} \left[ \Sigma_\lambda({\bm 0},\omega)\right]$, calculated from Eq.~(\ref{eq:imsigma}), is plotted as a function of $\omega$ in Fig.~\ref{fig:two}(b), for $\varepsilon_{\rm F} = 400~{\rm meV}$ and $d =60~{\rm nm}$.
(The dependence of the decay rate on $\varepsilon_{\rm F}$ and $d$ is discussed in Sect.~III of Ref.~\onlinecite{SMF}.)
We clearly see several peaks in ${\rm Im} \left[ \Sigma_\lambda({\bm 0},\omega)\right]$ for $\hbar \omega<  - \varepsilon_{\rm F}$ ($E_{\rm i} <0$), which occur at values of $\hbar\omega$ that are in a one-to-one correspondence with the ``resonant'' plasmon-phonon polaritons, i.e.~polaritons with group velocity equal to $v_{\rm F}$, shown in Fig.~\ref{fig:two}(a).
Indeed, as stated above, peaks in ${\rm Im} \left[ \Sigma_\lambda({\bm 0},\omega)\right]$ are expected at values of $\hbar\omega$---marked by green vertical lines in Fig.~\ref{fig:two}(b)---given by $\hbar \omega = \hbar v_{\rm F} q^\star - \varepsilon_{\rm F}  - \hbar \Omega_{{\bm q}^\star}$, where $q^\star$ is the wave number at which the resonance condition $d\Omega_{\bm q}/dq = v_{\rm F}$ is satisfied.
For example, the resonant mode at highest energy in Fig.~\ref{fig:two}(a), which occurs at $q^\star \approx 0.26~{\rm nm}^{-1}$ and energy $\hbar\Omega_{{\bm q}^\star} \approx 0.36~{\rm eV}$, yields a peak in ${\rm Im} \left[ \Sigma_\lambda({\bm 0},\omega)\right]$ at $\hbar\omega/\varepsilon_{\rm F} \approx -1.5$, see Fig.~\ref{fig:two}(b).

Comparing Fig.~\ref{fig:two}(b) with Fig.~\ref{fig:two}(c), we clearly see the role of dynamical screening due to electron-electron interactions in graphene.
For $\varepsilon({\bm k}, \omega) = 1$, the off-shell decay rate ${\rm Im} \left[ \Sigma_\lambda({\bm 0},\omega)\right]$ shows only a polaron peak, due to the emission of a Fabry-Per\'ot PP mode with group velocity equal to $v_{\rm F}$, see Fig.~1 in Ref.~\onlinecite{SMF}.

At ${\bm k} \neq {\bm 0}$, the conduction and valence band ${\rm Im} [\Sigma_{\lambda}({\bm k},\omega)]$ plasmon-phonon polariton peaks broaden and separate, because of~\cite{polini_prb_2008} the dependence on scattering angle of $\xi_{\lambda', {\bm k}+{\bm q}}$ and the chirality factor ${\cal F}_{\lambda\lambda'}$, which emphasizes ${\bm k}$ and ${\bm q}$ in nearly parallel directions for conduction band states and ${\bm k}$ and ${\bm q}$ in nearly opposite directions for valence band states.
As a result, the conduction band plasmon-phonon polariton peak moves up in energy while the valence band peak moves down. 

{\it Quasiparticle spectral function.---}An ARPES experiment~\cite{damascelli_rmp_2003} probes the quasiparticle spectral function ${\cal A}({\bm k},\omega) = -\pi^{-1} \sum_{\lambda} {\rm Im} [G_{\lambda}({\bm k}, \omega)] = \sum_{\lambda = \pm 1}{\cal A}_\lambda({\bm k},\omega)$ of the occupied states below the Fermi energy.
Here $G_{\lambda}({\bm k}, \omega)$ is the one-body Green's function in the band representation and 
\begin{equation}\label{eq:spectralfunction}
{\cal A}_\lambda =  - \frac{1}{\pi}\frac{{\rm Im} \Sigma_\lambda}{(\omega - \xi_{\lambda, {\bm k}}/\hbar - {\rm Re} \Sigma_\lambda/\hbar)^2 + ({\rm Im} \Sigma_\lambda/\hbar)^2}~.
\end{equation}
In writing Eq.~(\ref{eq:spectralfunction}) we have dropped explicit reference to the ${\bm k},\omega$ variables.
The real part ${\rm Re} [\Sigma_\lambda({\bm k}, \omega)]$ of the quasiparticle self-energy can be calculated, at least in principle, from the Kramers-Kronig transform of ${\rm Im} [\Sigma_\lambda({\bm k}, \omega)]$.
A more convenient way to handle the numerical evaluation of ${\rm Re} [\Sigma_\lambda({\bm k}, \omega)]$ is to employ the Quinn-Ferrell line-residue decomposition~\cite{quinn_pr_1958}.

Our main results for the quasiparticle spectral function ${\cal A}({\bm k}, \omega)$ of a doped graphene sheet placed on a hBN slab are summarized in Fig.~\ref{fig:one} and Fig.~\ref{fig:three}.
We clearly see that the presence of the hBN substrate is responsible for the appearance of a family of sharp dispersive satellite features associated with the presence of PPs and plasmon-phonon polaritons.
This is particularly clear in the one-dimensional cut at ${\bm k} = {\bm 0}$ of ${\cal A}({\bm k}, \omega)$ displayed in Fig.~\ref{fig:three}(a) for $d=10~{\rm nm}$.
All the sharp structures between the ordinary quasiparticle peak slightly below $\hbar \omega = - 0.4~{\rm eV}$ and the peak at $\hbar \omega \approx -0.7~{\rm eV}$, which is mostly plasmonic in nature, are sensitive to the detailed distribution and dispersion of Fabry-Per\'ot PP in the hBN slab, and therefore to the slab thickness $d$.
This is clearly shown in Fig.~\ref{fig:three}(b) and~(c), where we see shifts of these peaks of several  ${\rm meV}$, when $d$ is changed from $d = 10~{\rm nm}$ to $d = 60~{\rm nm}$, while keeping $\varepsilon_{\rm F}$ constant.

In summary, we have studied the coupling between standing phonon-polariton modes in a hyperbolic crystal slab and the plasmons of the two-dimensional massless Dirac fermion liquid in a nearby graphene sheet.
We have shown that this coupling yields a complex spectrum of (plasmon-phonon) polaritons, see Fig.~\ref{fig:two}(a).
Plasmon-phonon polaritons with group velocity equal to the graphene Fermi velocity couple strongly with graphene quasiparticles, enabling ARPES access to PP modes in hyperbolic crystal slabs, as shown in Figs.~\ref{fig:one} and~\ref{fig:three}.
Recent progress~\cite{roth_nanolett_2013} in the chemical vapor deposition 
growth of large-area graphene/hBN stacks on Cu(111) in ultrahigh vacuum and the ARPES characterization of the resulting samples makes us very confident on the observability of our predictions.
Our findings suggest that appropriate coupling of graphene to substrates which allow strong plasmon-phonon hybridization could open the route to the manipulation of carriers' spectral properties, paving the way for novel device functionalities.

\noindent {\it Acknowledgements.---}We gratefully acknowledge F.H.L. Koppens for useful discussions.
This work was supported by the EC under the Graphene Flagship program (contract no.~CNECT-ICT-604391) (A.T. and M.P.), 
MIUR (A.T. and M.P.) through the programs ``FIRB - Futuro in Ricerca 2010'' - Project ``PLASMOGRAPH'' (Grant No.~RBFR10M5BT) and ``Progetti Premiali 2012'' - Project ``ABNANOTECH'', 
the U.S.~Department of Energy under grant DE-FG02-05ER46203 (A.P.), and a Research Board Grant at the University of Missouri (A.P.).
Work at MIT was supported as part of the Center for Excitonics, an Energy Frontier Research Center funded by the U.S.~Department of Energy, Office of Science, Basic Energy Sciences under Award No.~desc0001088.
This work was also supported, in part, by the U.S.~Army Research Laboratory and the U.S.~Army Research Office through the Institute for Soldier Nanotechnologies, under contract number W911NF-13-D-0001.
Free software (www.gnu.org, www.python.org) was used.

\clearpage


\section*{Supplementary material}

# Supplemental Material for "Accessing phonon polaritons in hyperbolic crystals by ARPES"

Andrea Tomadin,[1] Alessandro Principi,[2] Justin C.W. Song,[3] Leonid S. Levitov,[4, *] and Marco Polini[1, 5, †]

[1]*NEST, Istituto Nanoscienze-CNR and Scuola Normale Superiore, I-56126 Pisa, Italy*
[2]*Department of Physics and Astronomy, University of Missouri, Columbia, Missouri 65211, USA*
[3]*Walter Burke Institute for Theoretical Physics and Institute for Quantum Information and Matter, California Institute of Technology, Pasadena, CA 91125, USA*
[4]*Department of Physics, Massachusetts Institute of Technology, Cambridge, Massachusetts 02139, USA*
[5]*Istituto Italiano di Tecnologia, Graphene Labs, Via Morego 30, I-16163 Genova, Italy*

In this Supplemental Material File we present i) analytical details on the Coulomb interaction potential dressed by the presence of an anisotropic dielectric, ii) numerical and analytical results on the dispersion of the phonon-polariton modes, and iii) numerical results on the quasiparticle decay rates and spectral function as the Fermi energy and the dielectric thickness are varied.

## I. ELECTROSTATICS AND FREQUENCY-DEPENDENCE OF THE HBN DIELECTRIC TENSOR

We consider a vertical heterostructure [inset of Fig. 2(b) in the main text] composed of a graphene sheet located at $z = 0$ and placed over a homogeneous but *anisotropic* insulator of thickness $d$ with dielectric tensor $\hat{\epsilon} = \mathrm{diag}(\epsilon_x, \epsilon_y, \epsilon_z)$. Homogeneous and isotropic insulators with dielectric constants $\epsilon_\mathrm{a}$ and $\epsilon_\mathrm{b}$ fill the two half-spaces $z > 0$ and $z < -d$, respectively. We calculate the electrical potential created by an electron in graphene for arbitary values of $\epsilon_x \neq \epsilon_y$ and then specialize our result to the case of a uniaxial crystal with $\epsilon_x = \epsilon_y$, such as hBN.

The electrical potential can be calculated from the following elementary approach. The displacement field $\boldsymbol{D}(\boldsymbol{r}, z)$ must satisfy the condition $\boldsymbol{\nabla} \cdot \boldsymbol{D}(\boldsymbol{r}, z) = 0$ everywhere in space. However, the presence of an electron with charge density $-e\delta^2(\boldsymbol{r})\delta(z)$ at $z = 0$ implies a discontinuity of the normal component $D_z$ of the displacement field across $z = 0$, while the tangential components $E_x, E_y$ of the electric field $\boldsymbol{E}(\boldsymbol{r}, z)$ must be continuous.

Since the electric field $\boldsymbol{E}(\boldsymbol{r}, z)$ is irrotational everywhere in space, we can introduce the electric potential $V(\boldsymbol{r}, z)$ in the three regions of space $z > 0$, $-d < z < 0$, and $z < -d$. The Laplace equation $-\epsilon_x \partial_x^2 V(\boldsymbol{r}, z) - \epsilon_y \partial_y^2 V(\boldsymbol{r}, z) - \epsilon_z \partial_z^2 V(\boldsymbol{r}, z) = 0$ in the anisotropic dielectric (i.e. for $-d < z < 0$) can be reduced [1] to an ordinary Laplace equation by scaling $x \to x/\sqrt{\epsilon_x}$, $y \to y/\sqrt{\epsilon_y}$, and $z \to z/\sqrt{\epsilon_z}$. Imposing the aforementioned boundary conditions and carrying out elementary algebraic steps, we find the 2D Fourier transform $V_{\boldsymbol{q}}$ of the electrical potential:

$$V_{\boldsymbol{q}} = \varphi_{\boldsymbol{q}} \frac{\epsilon_z q \kappa + \epsilon_\mathrm{b} q^2 \tanh(d\kappa)}{\epsilon_z q \kappa + (\epsilon_\mathrm{a}\epsilon_\mathrm{b} q^2 + \epsilon_z^2 \kappa^2) \tanh(d\kappa)/(2\bar{\epsilon})} \ . \quad (1)$$

Here, $q = |\boldsymbol{q}|$, $\boldsymbol{q} = (q_x, q_y)$, $\kappa \equiv (\epsilon_x q_x^2/\epsilon_z + \epsilon_y q_y^2/\epsilon_z)^{1/2}$, and $\varphi_{\boldsymbol{q}} \equiv -2\pi e/(q\bar{\epsilon})$ with $\bar{\epsilon} = (\epsilon_\mathrm{a} + \epsilon_\mathrm{b})/2$. Eq. (1) is the most important result of this Section and reproduces all known elementary results. For example, in the

|  | $\ell = x$ | $\ell = z$ |
| --- | --- | --- |
| $\epsilon_{\ell,0}$ | 6.41 | 3.0 |
| $\epsilon_{\ell,\infty}$ | 4.54 | 2.5 |
| $\gamma_\ell$ (meV) | 0.82 | 0.23 |
| $\hbar\omega_\ell^\mathrm{T}$ (meV) | 168.0 | 94.2 |
| $\hbar\omega_\ell^\mathrm{L}$ (meV) | 199.6 | 103.2 |

TABLE I. Microscopic parameters [4] entering Eq. (3).

limit $d \to 0$ Eq. (1) yields the potential of an electron in a graphene sheet embedded in two homogeneous and isotropic dielectrics, i.e. $V_{\boldsymbol{q}} \to \varphi_{\boldsymbol{q}}$. Similarly, for any finite $d$ and for an isotropic medium with $\epsilon_x = \epsilon_y = \epsilon_z \equiv \epsilon_\mathrm{i}$, it is easy to check that Eq. (1) yields [2] $V_{\boldsymbol{q}} \to -4\pi e \cosh(\eta_\mathrm{a}) \cosh(qd + \eta_\mathrm{b})/[q\epsilon_\mathrm{i} \sinh(qd + \eta_\mathrm{a} + \eta_\mathrm{b})]$ with $2\eta_\mathrm{a,b} = \ln[(\epsilon_\mathrm{i} + \epsilon_\mathrm{a,b})/(\epsilon_\mathrm{i} - \epsilon_\mathrm{a,b})]$. The ratio on the right-hand side of Eq. (1) can be interpreted as a form factor due to the presence of the anisotropic insulator slab, which dresses the "bare" 2D Coulomb potential $\varphi_{\boldsymbol{q}}$. For graphene on hBN, the relevant limit of Eq. (1) is that of a *uniaxial* dielectric medium with $\epsilon_y = \epsilon_x$. In this case Eq. (1) reduces to

$$V_{\boldsymbol{q},\omega} = \varphi_{\boldsymbol{q}} \frac{\sqrt{\epsilon_x \epsilon_z} + \epsilon_\mathrm{b} \tanh(qd\sqrt{\epsilon_x/\epsilon_z})}{\sqrt{\epsilon_x \epsilon_z} + (\epsilon_x \epsilon_z + \epsilon_\mathrm{b} \epsilon_\mathrm{a}) \tanh(qd\sqrt{\epsilon_x/\epsilon_z})/(2\bar{\epsilon})} \ . \quad (2)$$

We have modified the notation of the Coulomb potential in Eq. (2) to explicitly indicate its dependence on frequency. Indeed, in the case of hBN, the components of the dielectric tensor have an important dependence on frequency in the mid infrared, which is usually parametrized in the following form [3]

$$\epsilon_\ell(\omega) = \epsilon_{\ell,\infty} + \frac{\epsilon_{\ell,0} - \epsilon_{\ell,\infty}}{1 - (\omega/\omega_\ell^\mathrm{T})^2 - i\gamma_\ell \hbar\omega/(\hbar\omega_\ell^\mathrm{T})^2} \ , \quad (3)$$

with $\ell = x$ or $z$. Here $\epsilon_{\ell,0}$ and $\epsilon_{\ell,\infty}$ are the static and high-frequency dielectric constants, respectively, while $\omega_\ell^\mathrm{T}$ is the transverse optical phonon frequency in the direction $\ell$. The longitudinal optical phonon frequency $\omega_\ell^\mathrm{L}$ satisfies the Lyddane-Sachs-Teller relation $\omega_\ell^\mathrm{L} =$

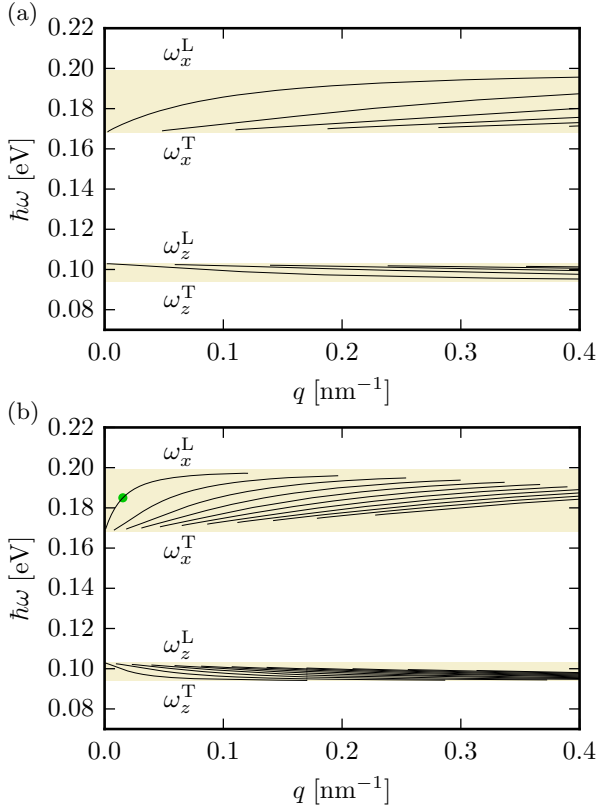

FIG. 1. (Color online) Dispersion of phonon-polaritons modes in a hBN slab. Poles of the dressed electrical potential $V_{\bm{q},\omega}$ in Eq. (2) are shown as functions of the wave vector $\bm{q}$. Here and in the following figures we use $\epsilon_{\rm a} = 1$ (vacuum) and $\epsilon_{\rm b} = 3.9$ (SiO$_2$). Panel (a) is for $d = 10$ nm, while panel (b) is for $d = 60$ nm. Shaded areas denote upper and lower reststrahlen bands. The green filled circle represents the point where the group velocity of the mode equals the graphene Fermi velocity $v_{\rm F}$. The 10 nm-thick slab does not support modes with group velocity equal to $v_{\rm F}$.

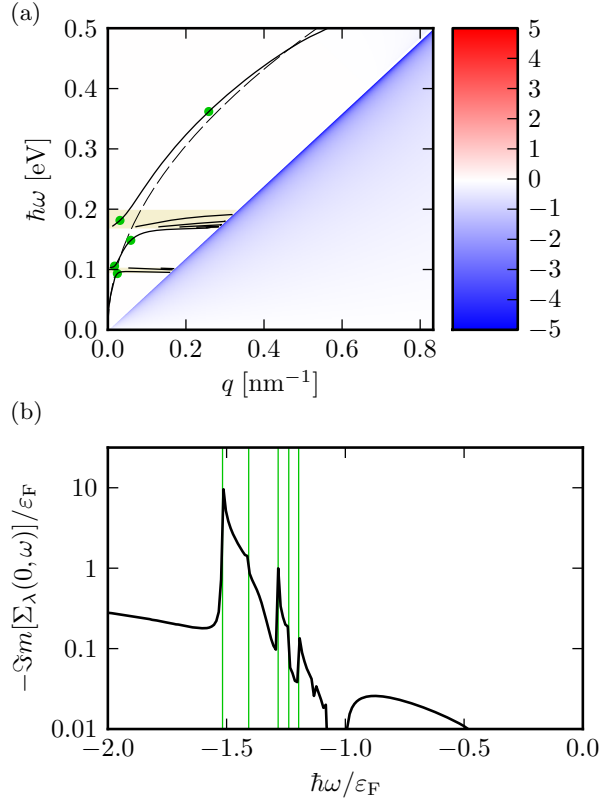

FIG. 2. (Color online) Panel (a) Plasmon-phonon polariton dispersion for a doped graphene sheet ($\varepsilon_{\rm F} = 400$ meV) on a 10 nm-thick hBN slab. The dashed line represents the dispersion relation of a Dirac plasmon in graphene in the absence of hBN phonons. Horizontal shaded areas and green filled circles have the same meaning as in Fig. 1. The density plot shows the imaginary part of the non-interacting polarization function in graphene and the corresponding colorbar is in units of the density of states at the Fermi energy. Panel (b) The quantity $-{\rm Im}[\Sigma_\lambda(\bm{k},\omega)]$ (in units of $\varepsilon_{\rm F}$ and evaluated at $\bm{k} = 0$) is shown as a function of the rescaled frequency $\hbar\omega/\varepsilon_{\rm F}$. Green vertical lines denote the values of $\hbar\omega/\varepsilon_{\rm F}$ at which a plasmon-phonon polariton peak is expected.

$\omega_\ell^{\rm T}\sqrt{\epsilon_{\ell,0}/\epsilon_{\ell,\infty}}$. The parameters in Eq. (3) are listed in Table I and have been taken from recent measurements [4] on high-quality *bulk* hBN [5]. A simple inspection of Eq. (3) in the limit $\gamma_\ell \to 0$ shows that, as we stated earlier, hBN is a hyperbolic material: in the lower (upper) reststrahlen band, which is defined by the inequality $\omega_z^{\rm T} < \omega < \omega_z^{\rm L}$ ($\omega_x^{\rm T} < \omega < \omega_x^{\rm L}$), the quantities $\epsilon_x\epsilon_z, \epsilon_x/\epsilon_z$ take negative values.

## II. ANALYTICAL RESULTS ON THE DISPERSION PHONON-POLARITON MODES

In the limit $\gamma_\ell \to 0$ it is possible to obtain analytical expressions for the dispersion of phonon-polariton modes in a hBN slab (without a graphene sheet placed on top of it). We remind the reader that these modes are the solutions $\omega = \omega_{\bm{q}}$ of the equation $\sqrt{|\epsilon_x(\omega)\epsilon_z(\omega)|} + (2\bar\epsilon)^{-1}[\epsilon_x(\omega)\epsilon_z(\omega) + \epsilon_{\rm b}\epsilon_{\rm a}]\tan[qd\sqrt{\epsilon_x(\omega)/\epsilon_z(\omega)}] = 0$.

Here, we report the results for the modes in the upper reststrahlen band. Similar expressions can be obtained for the lower reststrahlen band. For $qd \ll 1$, we find that one mode approaches the bottom of the reststrahlen band linearly

$$\omega_{\bm{q}} \to \omega_x^{\rm T}\left(1 + \frac{\epsilon_{x,0} - \epsilon_{x,\infty}}{4\bar\epsilon}\,qd + \dots\right) \quad (4)$$

while the other modes decrease quadratically

$$\omega_{\bm{q}} \to \omega_x^{\rm T}\left[1 + \frac{1}{2}\frac{\epsilon_{x,0} - \epsilon_{x,\infty}}{\epsilon_z(\omega_x^{\rm T})}\left(\frac{qd}{n\pi}\right)^2 + \dots\right]. \quad (5)$$

In the limit $qd \ll 1$ (with $n = 1, 2, \dots$), instead we find that all modes approach the upper end of the reststrahlen

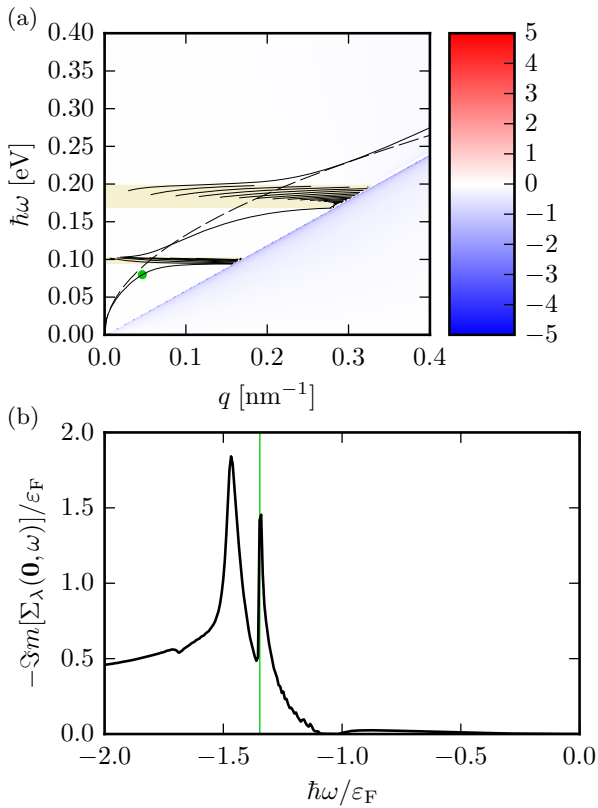

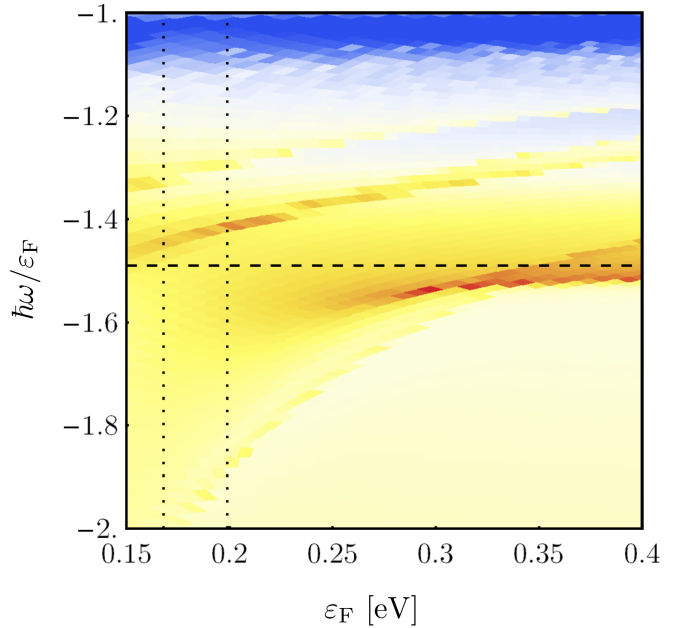

FIG. 3. (Color online) Same as Fig. 2 with parameters $\varepsilon_{\rm F} = 150$ meV and $d = 60$ nm.

band asymptotically

$$\omega_{\bm q} \to \omega_x^{\rm L} \left\{ 1 - \frac{1}{2}\epsilon_z(\omega_x^{\rm L}) \frac{\epsilon_{x,0}-\epsilon_{x,\infty}}{\epsilon_{x,\infty}\epsilon_{x,0}} \left(\frac{n\pi}{qd}\right)^2 + \ldots \right\}. \quad (6)$$

## III. ADDITIONAL NUMERICAL RESULTS

The following figures contain additional numerical results with respect to those included in the main text.

Figs. 2 and 3 show the plasmon-phonon polariton mode dispersion and the frequency dependence of the imaginary part of the quasiparticle self-energy. With respect to Figs. 2(a) and 2(b) in the main text, different values of hBN thickness and graphene Fermi energy are used in Fig. 2 and 3, respectively. In Fig. 2, we see that the dispersion of the plasmon-phonon polaritons changes with the hBN slab thicknes, and so does the wave vector where the group velocity of the modes equals the graphene Fermi velocity $v_{\rm F}$. However, the connection between the modes slope and the peaks in the imaginary part of the quasiparticle self-energy persists. In Fig. 3, the lower chemical potential induces a less steep dispersion of the bare plasmon mode. This reflects in the fact that all the plasmon-phonon polariton modes are less steep and the number of points where the mode group velocity matches $v_{\rm F}$ is reduced from 7 [cfr. Figs. 2(a) of the main text] to 1. However, a large segment of the hybrid mode branch has *approximately* group velocity equal to $v_{\rm F}$ in the range $0.1~{\rm nm}^{-1} \lesssim q \lesssim 0.3~{\rm nm}^{-1}$ and $0.10$ eV $\lesssim \hbar\omega \lesssim 0.15$ eV. This results in an enhanced decay rate which appears as a broad peak of the imaginary part of the self-energy at $\hbar\omega/\varepsilon_{\rm F} \simeq 1.5$.

The dependence of the decay rate (proportional to the imaginary part of the quasiparticle self-energy) on the Fermi energy is studied in detail in Fig. 4. We see that the decay rate depends smoothly on the Fermi energy and that there is no abrupt change in the imaginary part of the quasiparticle self-energy when the Fermi energy crosses the extremes of the upper reststrahlen band. Moreover, for large Fermi energies, the main peak in the imaginary part of the self-energy is due to a plasmon-polariton mode which exists also in the absence of the hBN. This mode corresponds to the higher-energy branch in Fig. 2(a) of the main text and in Fig. 2 and generates the spectral feature labeled by (2) in Fig. 1 of the main text. To see this, in Fig. 4 we plot (horizontal dashed line) the expected position of the plasmon-polariton peak in the absence of the hBN slab. This is easily found from the well-known dispersion of the graphene plasmon, which has group velocity equal to $v_{\rm F}$ at the wave

FIG. 4. (Color online) A 2D color plot of the quantity $-{\rm Im}[\Sigma_\lambda(\mathbf{0},\omega)]$ as a function of $\hbar\omega/\varepsilon_{\rm F}$ and $\varepsilon_{\rm F}$. Data in this plot refer to a hBN slab of thickness $d = 60$ nm. As usual, $\epsilon_{\rm a} = 1$ and $\epsilon_{\rm b} = 3.9$. The horizontal dashed line marks the expected position of a plasmon-phonon polariton peak in the absence of the hBN slab. The vertical dotted lines mark the extremes of the upper reststrahlen band.

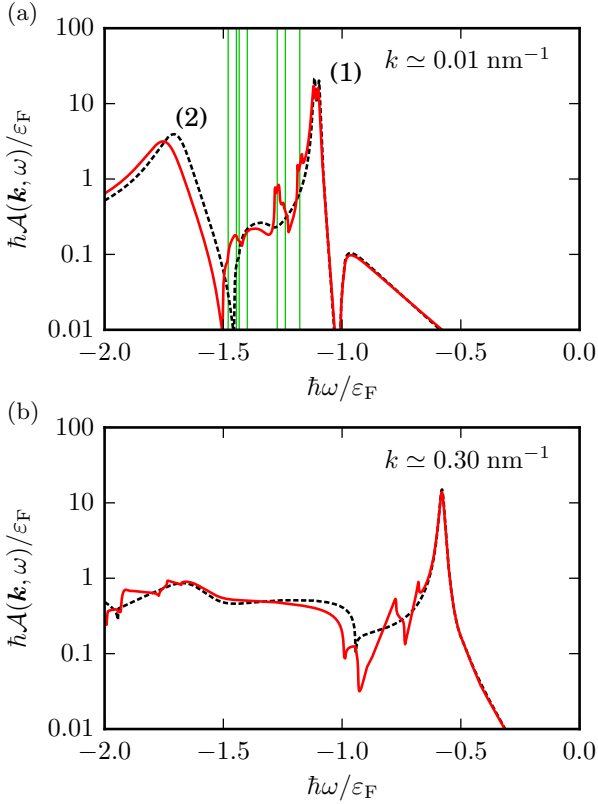

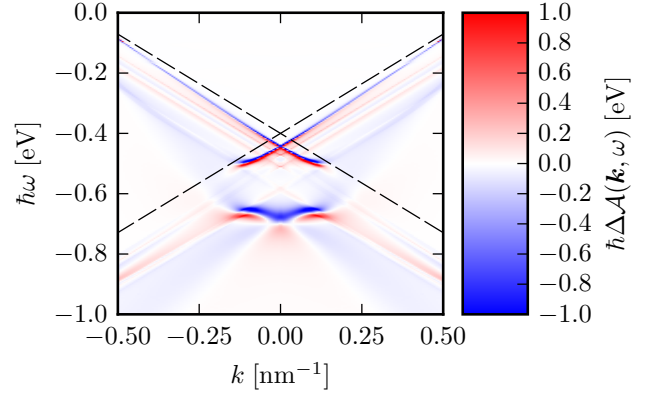

FIG. 6. (Color online) Difference between the spectral function $\mathcal{A}(\bm{k},\omega)$ for $d = 60$ nm and the spectral function calculated in the absence of the hBN slab, i.e. $d = 0$. The Fermi energy is $\varepsilon_{\mathrm{F}} = 400$ meV. The dashed line indicates the Dirac cone. The redistribution of spectral weight due to the presence of the hBN slab is sizable.

FIG. 5. (Color online) Frequency dependence of the quasi-particle spectral function $\mathcal{A}(\bm{k},\omega)$ for two values of $k$ (indicated in the plots), corresponding to $k = 0.01$ and $0.5\,k_{\mathrm{F}}$. The Fermi energy is $\varepsilon_{\mathrm{F}} = 400$ meV. The red solid lines correspond to $d = 60$ nm, while the black dashed line correspond to $d = 0$. Panel (a) The vertical green lines correspond to the expected positions of the plasmon-phonon polariton peaks for $k \simeq 0$. The two spectral features labeled (1) and (2) correspond to those in Fig. 1 of the main text.

vector $q^{\ast} = \varepsilon_{\mathrm{F}} e^{2}/(\epsilon_{\mathrm{a}} + \epsilon_{\mathrm{b}})$. Then, following the argument explained in the main text, we find that the peak in the imaginary part of the quasiparticle self-energy is expected at $\hbar\omega/\varepsilon_{\mathrm{F}} = -e^{2}/(\hbar v_{\mathrm{F}}\bar{\epsilon}) + \hbar v_{\mathrm{F}} q_{\parallel} - 1 \simeq -1.5$, independent of other parameters. Satellite modes are clearly visible and drift to lower values of $\hbar\omega$ with decreasing Fermi energy. The main peak broadens as the upper reststrahlen band is approached, and eventually a satellite peak becomes more prominent and closer to the frequency $\hbar\omega/\varepsilon_{\mathrm{F}} \simeq -1.5$ of the main plasmon-polariton mode.

Fig. 5 shows the frequency dependence of the quasi-particle spectral function at finite wave vector. Each panel shows the spectral function with (red solid line) and without (black dashed line) the hBN substrate. In Fig. 5(a), the quasiparticle peaks corresponding to conduction and valence band, labeled by (1), are very close in energy and barely resolved. The spectral feature at $\hbar\omega/\varepsilon_{\mathrm{F}} \simeq -1.7$, labeled by (2), is due to the plasmon-polariton mode discussed above. The satellite peaks between (1) and (2) appear close to the frequencies where the peaks of the imaginary part of the quasiparticle self-energy are present. However, the expected position of the plasmon-polariton peak does not match its actual position (2). This effect is well-known and is a consequence of the attractive interaction between the emitted plasmon and the hole left behind by the photo-excited electron. [7] For the satellite peaks this effect is smaller as the frequency $\omega$ is increased. This can be understood from the plot of the imaginary part of the self-energy [Fig. 2(b) in the main text]: the plasmon-polariton peak has the largest decay rate, which obviously changes the position where the spectral function is maximal. The decay rate of the satellite peaks decreases as the frequency $\omega$ is increased. Inspecting Fig. 1 of the main text, we also note that $k \simeq 0$ or $k \lesssim k_{\mathrm{F}}$ are the wave vector ranges where the satellite peaks could be more clearly seen, because at intermediate values of $k$ they are obfuscated as they cross the broad spectral features (1) and (2). Fig. 5(b) shows that at intermediate values of $k$ the peak corresponding to the conduction band is clearly seen, while the valence band peak is strongly damped and disappears in the continuum of excitations for $\hbar\omega \lesssim \varepsilon_{\mathrm{F}}$.

Finally, Fig. 6 gives a more complete representation of the data shown in Fig. 5, by emphasizing the difference between the quasiparticle spectral function with and without the hBN substrate. The effect of the substrate is sizable, with variations in magnitude of the spectral function of order eV/$\hbar$. The spectral weight of both peaks (1) and (2) moves slightly to lower energies (i.e. the blue regions "move" to the red regions when $d$ has a finite thickness). Moreover, the fine satellite bands are very clearly seen as an alternation of red and blue regions in this plot.

* levitov@mit.edu
† marco.polini@icloud.com



\end{document}